\documentstyle[12pt]{article}
\begin{document}
\hspace*{8cm}{\large JINR preprint E2-94-488}

\vspace{1cm}

\begin{center}
{\large \bf TO THE PROBLEM OF $1/N_c$ APPROXIMATION IN THE NAMBU-
JONA-LASINIO MODEL} \\

\vspace{1.5cm}
{\large

D.Ebert

\vspace{0.5cm}
{\em Institut f\"ur Elementarteilchenphysik,
Humboldt-Universit\"at, \\
Invaliden Str. 110, D-10115 Berlin, FRG} \\

\vspace{1.0cm}

M.Nagy

\vspace{0.5cm}
{\em Institute of Physics of Slovak Academy of Sciences, \\
842 28 Bratislava, Slovakia} \\

\vspace{1.0cm}

M.K.Volkov

\vspace{0.5cm}
{\em Joint Institute for Nuclear Research, \\
141980 Dubna, Russian Federation}   }\\

\vspace{2.0cm}

\end{center}
\vspace{2cm}

\begin{center}

{\large\bf Abstract}
\end{center}

In this article, the gap equation for the constituent quark mass
in the U(2)$\times$U(2) Nambu-Jona-Lasinio model for the $1/N_c$
approximation is investigated. It is shown that taking into
account scalar isovector mesons plays an important role for
the correct description of quark masses in this approximation.
The role of the Ward identity in calculations of $1/N_c$ corrections
to the meson vertex functions is shortly discussed.

\newpage

The NJL model in the leading $1/N_c$ approximation, Hartree
approximation, allows us to obtain
a relatively complete picture of low-energy meson physics [1-5]
($N_c$ is the number of quark colors).
However, in the last time, there have been undertaken some
attempts to describe next to the leading $1/N_c$ approximation in
the NJL model and to consider mesons not in the tree diagrams only
but also in the loop diagrams [6-12]. Interesting results have
been obtained in this direction for the description of the behaviour
of the thermodynamical potential and the bulk of thermodynamical
quantities in the vicinity of the critical temperature. It has
been shown that mesonic degrees of freedom play the dominant
role at low $T$, and the quark degrees of freedom are most
relevant at high $T$.

Thus, it seems to be very useful to continue these investigations,
to study more carefully the $1/N_c$ approximation in the NJL
model by using different methods. Here, we consider the perturbation
theory and calculate $1/N_c$ corrections to the gap equation.
We will show how to correctly use the
perturbation theory for the description of constituent quark mass
in the $1/N_c$ approximation. Our results are remarkably different from
those obtained in the series of previous papers (see e.g. [7]). It
will be shown that the inclusion of scalar isovector mesons
$a_0(980)$ plays an important role in the description of the $1/N_c$
approximation.

We consider the NJL model for the $U(2)\times U(2)$ chirally symmetric
case [1-2]

$$
L({\bar q},q)={\bar q}(i{\hat \partial}- m^0)q + {G\over 2}[({\bar
q}{\lambda}^a q)^2 + ({\bar q}i\gamma_5{\lambda}^aq)^2],    \eqno(1)
$$

\noindent where $q$ are the fields of u and d quarks, $m^0$ is the
current quark mass, $\lambda^0 =1$ is the unique matrix and
$\lambda^a$ = $\tau^a$ $(a=1,2,3)$ are the Pauli matrices.

After the introduction of meson fields by using the technique of
generating functional [1-3] and performing the integration over
quark fields in the functional integral, we come to the Lagrangian

$$
L'({\tilde \sigma},\phi)= -{{\tilde \sigma}^2_a + \phi^2_a \over
2G} - i{\rm Tr}\ln S^{-1}(x-y),    \eqno(2)
$$

\noindent where ${\tilde \sigma}_a$ and $\phi_a$ are the scalar
and pseudoscalar meson fields, respectively, ${\tilde \sigma}_0
= \sigma_0 - m + m_0~$, ${\tilde \sigma}_a = \sigma_a$ $(a=1,2,
3)$

$$
S^{-1}(x,y) = [i{\partial}_x - m + \sigma_a\lambda^a +
i\gamma_5\lambda^a\phi_a]{\partial}^4(x-y).     \eqno(3)
$$

To get the $\sigma$-model, it is enough to consider the divergent
quark loops, depicted in Fig.1, and to perform the renormalization
of meson fields [1-3]

As a result, we obtain the meson Lagrangian of the following type:

$$
L^{''}(\sigma , \phi)= {1\over 4} {\rm Tr} \left\{ (\partial_{\mu}
{\bar\sigma})^2 + (\partial_{\mu}{\bar\phi})^2 + 2g\left({m-m_0
\over G}-8mI_1(m, \Lambda)\right){\bar \sigma}- \right.
$$
$$
-\left.g^2\left({1\over G}-8I_1(m,\Lambda)\right)({\bar \sigma}^2 +
{\bar\phi}^2)-g^2\left[{\bar \sigma}^2 -2{m\over g}{\bar \sigma} +
{\bar \phi}^2 \right]^2\right\} -
$$
$$
-i{\rm Tr}\ln\left\{1+ {g\over i{\hat \partial}-m}[{\bar \sigma}+
i\gamma_5{\bar \phi}]\right\},  \eqno(4)
$$

\noindent where $g=[4I_2(m,\Lambda)]^{-1/2}$, ${\bar \sigma}=
\sigma^a\lambda_a$, ${\bar \phi} = \phi^a\lambda_a$, and
$I_1(m,\Lambda)$ and $I_2(m,\Lambda)$ are divergent integrals
($\Lambda$ is the cut-off parameter)
$$
I_n(m,\Lambda) = -i{N_c\over (2\pi)^4}\int^{\Lambda} {d^4k\over
(m^2-k^2)^n}.   \eqno(5)
$$

 From (5) we can see that the coupling constant $g^2$ has the order $1/N_c$.
Remind that the coupling constant $G$ also has the order $1/N_c$.

 From the condition ${\delta L''(\sigma,\phi)\over \delta
\sigma}\vert_{\sigma,~\phi=0}=0$ (absence of linear in $\sigma$
terms in $L''(\sigma, \phi)$) we obtain the gap equation

$$
m = m^0 + 8mGI_1(m, \Lambda)~.    \eqno(6)
$$

How does the gap equation change, if we permit the existence of
meson propagators inside the quark loops? ($1/N_c$ approximation).
 From Fig.1 one can easily see that in this case, in addition
to the tadpole 1a there appear complementary terms (linear in
$\sigma$) from the diagram 1c which lead to the appearance of
additional terms in the gap equation (6) (see Fig.2)

$$
m =m^0 + 8mGI_1(m, \Lambda) + \Delta =
$$
$$
=m^0 + 2G{iN_c\over (2\pi)^4}{\rm Tr}\int^{\Lambda} {d^4k\over
{\hat k}-m} + 2G{iN_c\over (2\pi)^4}{\rm Tr}\int^{\Lambda} d^4k{1\over
{\hat k}-m}\Sigma(k){1\over {\hat k}-m} + ...    \eqno(7)
$$

\noindent The last two terms in (7) can be written in the form of
one tadpole with the modified quark mass:

$$
m = m^0 + 2G{iN_c\over (2\pi)^4}{\rm Tr}\int^{\Lambda} {d^4k\over {\hat k}- m
-\Sigma(k)}~,    \eqno(7a)
$$

\noindent where $\Sigma(k)$ is the operator of quark self-energy

$$
\Sigma(k) = 3\Sigma_{\pi}(k) + \Sigma_{\sigma_0}(k) +
3\Sigma_{a_0}(k)~,    \eqno(8)
$$
$$
\Sigma_{\pi}(k) = i{g^2_{\pi}\over (2\pi)^4}\int^{\bar{\Lambda}} d^4q {{\hat
q}-m
\over (m^2-q^2)(M^2_{\pi}-(k-q)^2)},    \eqno(9)
$$
$$
\Sigma_{\sigma_i}(k) = i{g^2\over (2\pi)^4}\int^{\bar{\Lambda}} d^4q {{\hat
q}+m
\over (m^2-q^2)(M^2_{\sigma_i}-(k-q)^2)}.   \eqno(10)
$$

\noindent Here $M_{\pi}$ and $M_{\sigma_i}$ are the masses of
pions and $\sigma$-particles $(\sigma_i = \sigma_0, a^0_0, a_0^+,
a_0^-)$, respectively, $g_{\pi}={m\over F_{\pi}}$, where
$F_{\pi}=93$ MeV is the pion decay constant .
\footnote{\normalsize After accounting $\pi - a_1$ transitions
the constants $g_{\sigma}$ and $g_{\pi}$ will be different from
each other [1b]. Here $a_1$ is the axial vector meson.

In the general case the cut-off parameters $\Lambda$ and $\bar{\Lambda}$
are not equal to each other .Here, we assume that
$\Lambda$ = $\bar{\Lambda}$ = 1.2GeV.}

The gap equation (7a) can be written in the form of the
Schwinger - Dyson equation for the new quark mass ${\bar m}$
=$ m + \Sigma(m)$.
For this purpose, we add the term $\Sigma({\bar m})$ to both
the sides of equation (7a) and write it in the form

$$
{\bar m} = m^0 + 2G{iN_c\over (2\pi)^4}{\rm Tr}\int^{\Lambda} {d^4k\over
{\hat k}- {\bar m}} + \Sigma({\bar m})=
$$
$$
= m^0 + 8G{\bar m}I_1({\bar m},\Lambda) + \Sigma({\bar m})~.
\eqno(11)
$$

\noindent From equation (11) we can find the correction
$\delta m$ to the quark mass $m_H$, obtained in the Hartree
approximation, after taking account of the first order in $1/N_c$
expansion. That is why we write the mass $\bar m$ in the form

$$
{\bar m} = m_H + \delta m       \eqno(12)
$$

\noindent and expand the second term in the r.h.s. of (11) over $\delta m $,
conserving the terms of first order over $1/N_c$

$$
m_H + \delta m = m_0 +(m_H + \delta m)8G\left[I_1(m_H, \Lambda)
+\delta m {\delta I_1\over \delta m}\vert_{m=m_H}\right] +
\Sigma(m_H)~.    \eqno(13)
$$

\noindent By using the formulae

$$
{\delta I_1(m,\Lambda)\over \delta m} = -2mI_2(m,\Lambda) =
-{m\over 2g^2}     \eqno(14)
$$

\noindent and the gap equation in the Hartree approximation (see
formula (6))

$$
m_H=m_0 + 8Gm_HI_1(m_H,\Lambda),
$$

\noindent we find for $\delta m$ the following expression:

$$
\delta m = Z^{-1} \Sigma (m_H),  \eqno(15)
$$

\noindent where

$$
Z=16 Gm_H^2I_2(m_H,\Lambda) + {m_0\over m_H} = \left({2m_H\over
g}\right)^2 G +{m_0\over m_H}~.      \eqno(16)
$$

\noindent For the parameters we  use here [1b]:
$m_H = 280$ MeV, $m_0= 3.3$ MeV, $\Lambda = 1.2$ GeV, $G=5.4$
GeV$^{-2}$,and $g^2\approx 2\pi$, we get \footnote{\normalsize The
results, obtained in the papers [7,8], correspond to the
value $Z=1$.}

$$
Z^{-1}=3.6,~~~ \delta m = 3.6~\Sigma (m_H, \Lambda).
$$

Now we have to determine the operators $\Sigma_{\sigma_i}
(p,\Lambda)$ and $\Sigma_{\pi}(p,\Lambda)$ at the point ${\hat p}
=m_H$. One can easily evaluate the integrals in formulae (9) and (10)
and get the following expressions:

$$
\Sigma_{\pi}(p,\Lambda)= {g^2_{\pi}\over (4\pi)^2}\int_0^1 dx~
(m-x{\hat p})\left[\ln \left({\Lambda^2\over m^2}+1\right) +\right.
$$
$$
+\left.\ln{1+{\bar b_{\pi}}x +{\bar c}x^2\over 1+b_{\pi}x+cx^2}-
\left(1+{m^2\over\Lambda^2}\right)^{-1}{1\over 1+{\bar
b_{\pi}}x + {\bar c}x^2}\right]=
$$
$$
= {g_{\pi}^2\over (4\pi)^2}\left[ m C_1^{\pi}(p,\Lambda) - {\hat p}
C_2^{\pi}(p,\Lambda)\right ],   \eqno(17)
$$

$$
\Sigma_{\sigma_i}(p,\Lambda)=-{g^2\over (2\pi)^4}\int_0^1 dx~(m+
x{\hat p})\left[\ln \left({\Lambda^2\over m^2}+1\right) + \right.
$$
$$
+\left.\ln{1+{\bar b}_{\sigma_i}x + {\bar c}x^2\over
1+b_{\sigma_i}x+cx^2}- \left(1+{m^2\over
\Lambda^2}\right)^{-1}{1\over 1+{\bar b}_{\sigma_i}x+{\bar
c}x^2}\right]=
$$
$$
= -{g^2\over (4\pi)^2}\left[ m C_1^{\sigma_i}(p,\Lambda) + {\hat p}
C_2^{\sigma_i}(p,\Lambda)\right ],   \eqno(18)
$$

\noindent where

$$
b_i={M^2_i-m^2-p^2\over m^2},~~c={p^2\over m^2},~~{\bar b}_i=
{M^2_i-m^2-p^2\over a}, {\bar c}={p^2\over a}, a=m^2+\Lambda^2~,
$$
$$
C_1^i=\ln\left({\Lambda^2\over m^2} + 1\right)+ \left(1+{{\bar
b}_i\over 2{\bar c}}\right)\ln(1+{\bar b}_i+{\bar c})-\left(
2+{b_i\over c}\right)\ln{M_i\over m}+
$$
$$
+\left(1-{{\bar b}^2_i\over 2{\bar c}}\right){\bar I}_0+
\left({b^2_i\over 2c}-2\right)I_0~,    \eqno(19)
$$
\vskip0.3truecm
$$
C_2^i=-{1\over 2}\left({{\bar b}_i\over {\bar c}}- {b_i\over c}
\right)+ {1\over 2}\ln\left({\Lambda^2\over m^2} +1\right)+{1\over 2}
\left(1-{{\bar b}^2_i\over 2{\bar c}^2}\right)\ln(1+{\bar b}_i
+{\bar c})-
$$
$$
-\left[1+{1\over c}\left(1-{b^2_i\over 2c}\right)\right]
\ln{M_i\over m}- {{\bar b}_i\over 2{\bar c}}\left(1-{{\bar
b}^2_i\over 2{\bar c}}\right){\bar I}_0+ {b_i\over 2c}\left(
2-{b^2_i\over 2c}\right)I_0~,    \eqno(20)
$$
\vskip0.3truecm
$$
I_0=\int^1_0 {dx\over 1+b_ix+cx^2},~~~~~{\bar I}_0=\int^1_0
{dx\over 1+{\bar b}_ix+{\bar c}x^2}~.
$$

Scalars and pions give contributions to the quark mass with
the opposite signs and  strongly compensate each other. Therefore,
it is important to take into account contributions of all
mesons corresponding to the considered group of symmetry. In
our case of the group U(2)$\times$U(2), to the three pions there
correspond four scalar mesons in the scalar sector (scalar isoscalar
$\sigma_0(700)$ and three scalar isovectors $a_0(980)$).
\footnote{\normalsize The isoscalar partner of pions appears only
in the U(3)$\times$U(3) group in the form of a $\eta$ meson. Therefore,
we will not consider it here.Scalar mesons have the masses:
$m_{\sigma_0} = 700MeV$ and  $m_{a_0} = 980MeV$.}

\newpage

Table.1 gives the coefficients $C^i_1$ and
$C^i_2$ evaluated for all these mesons at $p^2=m^2$.

\vskip0.3truecm
\begin{table}[h]
\begin{center}
{\bf Table 1.}
\vskip0.3truecm
\begin{tabular}{ |c| c| c| c| c| c|}
\hline
$C_1^{\sigma_0}$  & ~$C_2^{\sigma_0}$  & ~$C_1^{a_0}$ & ~$C_2^{a_0}$ &
{}~$C_1^{\pi}$ & ~$C_2^{\pi}$  \\
\hline
{}~1.06  & ~0.42  & ~0.63  & ~0.5  & ~2.8  & ~1.5  \\
\hline
\end{tabular}
\end{center}
\end{table}
\vskip0.3truecm

Then, for $\Sigma(m,\Lambda)$ we obtain

$$
\Sigma(m,\Lambda)= {m\over (4\pi)^2}[-g^2(1.48 + 3\times 1.13)
+{g_{\pi}^2}(3\times 1.3) = -30.6 + 35.4 = 5]
$$
$$
\Sigma (m,\Lambda) = 0.03~,~~~~~~ \delta m = 0.11~m~.
$$

\noindent As a result, the mass of a constituent quark increases by
$11~\%$ and is equal to 310 MeV, which completely corresponds to the
standard value. If we consider only one scalar meson
$\sigma_0(700)$, the corrections rapidly increase, amounting to
$60~\%$ $(\delta m = 0.60~m)$, which does not correspond to the
$1/N_c$ approximation.
\footnote{\normalsize If we use the model values for the masses
of the scalar mesons:$m_{\sigma_i}^2 = m_{\pi}^2 + 4m^2$,
m = 580MeV, we get the negative value for $\delta m$. ($ C_1^{\sigma_i }$
= 1.3, $C_2^{\sigma_i }$ = 0.55 )}

These calculations have shown, that for the correct estimates
performed in the $1/N_c$ approximation, it is very important to take
into account all real contributions of mesons from the scalar
and pseudoscalar sectors.

As an example of another approach to estimation of the quark
mass in the $1/N_c$ approximation within the NJL model we can
illustrate the paper [7]. In this article two incorrect actions
have been done, in our opinion: the first
when the additional contributions (in the $1/N_c$ approximation) from
the leading tadpole term in the Schwinger-Dyson equation were not
taken into account. This led to the lowered result
which did not correspond to the real $1/N_c$ approximation.
At the second step, the contribution of only one
scalar isoscalar meson was considered instead of four
scalar mesons. This step substantially increased their estimate.
As a result of these two operations, the final $1/N_c$
corrections to the quark mass did not go beyond the limit of
$20~\%$ of the Hatree approximation.

One of the interesting tasks is the construction of chirally
symmetric perturbation theory for the $1/N_c$ expansion. The
positive results in this direction have been obtained by
G.S.Guralnik with coauthors already in 1976 [12].
They showed that in the $1/N_c$ approximation for the NJL model
with one scalar and one pseudoscalar mesons the pion mass was
equal to zero when the current quark mass was vanishing.
Therefore, the pion remains the Goldstone particle in this
approximation as well.

It is interesting to consider the changes of the meson coupling
constants $g$ and $g_{\pi}$ in the $1/N_c$ approximation. As we have
shown in the Appendix, the scalar meson coupling constant $g$
does not change in the $1/N_c$ approximation. A more complicated
situation took place for the coupling constants $g_{\pi}$ and
the Goldberger-Treiman identity.

 When this work has been fulfilled, we found out that a very
interesting paper appeared just now [13].In this work, a chirally
symmetric self-consistent $1/N_c$ approximation scheme to the NJL
model was developed.The authors used the correct $1/N_c$ approximation
for the gap equation and demonstrated explicitly that their scheme
fulfills all the chiral symmetry theorems - the Goldstone theorem,
Goldberger-Treiman relation and the conservation of the quark axial
current.

 This paper is very close to ref. [12]. In contrast with
our work they considered the $SU(2)\times SU(2)$ chiral symmetry Lagrangian
with only one scalar isoscalar meson and the case when the current quark mass
was equal to zero.

 In conclusion, we would like to say that the papers [12-13] and this one
give the full picture of the chirally symmetric $1/N_c$ approximation in
the NJL model.

 One of the authors (MKV) would like to express his gratitude to
Prof. J.H\"ufner and Dr. S.Klevansky for the useful discussions
and JSPS Program of Japan, INTAS fund (grant N 2915) and Russian
Fundation of the Fundamental Researches (grant N 93-02-14411) for
financial support.This work was supported also by DFG project 436 RUS 113.

\begin{center}
\section*{\bf APPENDIX}
\vskip0.2truecm
{\bf The scalar vertex function and Ward identity}
\end{center}

\vskip0.2truecm

Let us show that $1/N_c$ corrections to the scalar coupling
constant $g$ are equal to zero. For this aim we consider diagrams
depicted in Fig.3. The scalar vertex function for $\sigma_0$ meson
in $1/N_c$ approximation takes the form

$$
\Gamma^{(1/N_c)}(p,p'\vert q) = g_{\sigma} + \Gamma_{\sigma_0}^b
(p,p'\vert q) + \Gamma_{\sigma_0}^{(c+d)}(p,p'\vert q) +
3\Gamma_{a_0}^b(p,p'\vert q) + 3\Gamma_{a_0}^{(c+d)}(p,p'\vert q)+
$$
$$
+3 \Gamma_{\pi}^b(p,p'\vert q) +3\Gamma_{\pi}^{(c+d)}(p,p'\vert q).
\eqno(A.1)
$$

Now consider the case when $q=0$, $p=p'$. Then

$$
\Gamma_{\sigma_0}^{(b)}(p,p\vert 0) = -i{g^2\over (2\pi)^4}
\int {d^4k \over ({\hat k} + {\hat p}-m)^2(M^2_{\sigma_0}-k^2)},
\eqno(A.2)
$$
$$
\Sigma_{\sigma_0}(p+k) = -i{g^2\over (2\pi)^4}\int {d^4k \over
({\hat k} +{\hat p}-m)(M^2_{\sigma_0}-k^2)},   \eqno(A.3)
$$
$$
\Gamma^{(c+d)}_{\sigma_0}(p,p\vert 0)= {\Sigma_{\sigma_0}(p)-
\Sigma_{\sigma_0}(m)\over {\hat p} -m}\vert_{{\hat p}=m} =
{\delta \Sigma_{\sigma_0}(p)\over \delta {\hat p}}\vert_{{\hat p}=m}=
$$
$$
= i{g^2\over (2\pi)^4}\int {d^4k \over ({\hat k}+{\hat p}-m)^2
(M^2_{\sigma_0}-k^2)}= -\Gamma_{\sigma_0}^{(b)}(p,p\vert 0)
\eqno(A.4)
$$

The similar situation takes place for $\Gamma_{a_0}$ and
$\Gamma_{\pi}$. As a result all contributions of the diagrams
depicted in Fig.3b-d cancel each other and finally we got

$$
\Gamma^{(1/N_c)}(p,p\vert 0)= g_{\sigma}~.    \eqno(A.5)
$$

\eject


\begin{center}
\section*{\bf Figure captions}
\end{center}
\vskip1.0truecm

{\bf Fig.1} The quadratically (1a, 1b) and logarithmically (1c, 1d)
divergent quark loop diagrams in the NJL model.

\vskip1.0truecm

{\bf Fig.2} The additional tadpole diagram in the $1/N_c$ approximation.
The $\Sigma$ is the self-energy part of the quark propagator with
pion and scalar meson internal lines.

\vskip1.0truecm

{\bf Fig.3} The scalar vertex diagrams in the $1/N_c$ approximation .

\end{document}